\documentclass[10pt, conference]{IEEEtran}
\IEEEoverridecommandlockouts

\usepackage{cite}
\usepackage{amsmath,amssymb,amsfonts}
\usepackage{algorithmic}
\usepackage{graphicx}
\usepackage{textcomp}
\usepackage[table]{xcolor}
\usepackage[T1]{fontenc}
\usepackage{tikz}
\usepackage{multirow,multicol,ctable}
\usepackage{tabularx,booktabs}
\usepackage{diagbox}
\usepackage{xspace}
\usepackage{enumitem}
\usepackage{listings}
\usepackage[colorlinks=true,urlcolor=black,linkcolor=black,citecolor=black]{hyperref}
\usepackage{relsize}
\usepackage[capitalize]{cleveref} 
\usepackage{tikz}
\usepackage{calc}
\usepackage{txfonts}
\usepackage{microtype}

\newlength{\leftLength}

\newcommand{\summary}[2]{
\setlength{\leftLength}{0.5cm + (\widthof{{\small \textbf{#1}}}/2)}
\vspace{2ex}\noindent\begin{tikzpicture}
\node[align=center,draw,thin,minimum width=\columnwidth,inner sep=2.2mm] (titlebox)%
{\parbox{0.95\columnwidth}{\vspace*{0.25ex}\noindent\emph{#2}}};\node[fill=white] (W) at ([xshift=\the\leftLength] titlebox.north west) {{\small \textbf{#1}}};%
\end{tikzpicture}\vspace{2ex}}

\makeatletter

\usepackage{eso-pic}
\newcommand\AtPageUpperMyright[1]{\AtPageUpperLeft{
 \put(\LenToUnit{0.5\paperwidth},\LenToUnit{-1cm}){
     \parbox{0.5\textwidth}{\raggedleft\fontsize{9}{11}\selectfont #1}}
 }}
\newcommand{\conf}[1]{
\AddToShipoutPictureBG*{
\AtPageUpperMyright{#1}
}
}

\def\BibTeX{{\rm B\kern-.05em{\sc i\kern-.025em b}\kern-.08em
    T\kern-.1667em\lower.7ex\hbox{E}\kern-.125emX}}

\begin{document}

\title{Asset Management in Machine Learning: A Survey\\[.75ex]
}

\conf{\textit{International Conference on Software Engineering, track on Software Engineering in Practice (ICSE-SEIP) 23-29 May, 2021}} 

\author{\IEEEauthorblockN{Samuel Idowu\IEEEauthorrefmark{1}, Daniel Str\"{u}ber\IEEEauthorrefmark{2}, and Thorsten Berger\IEEEauthorrefmark{1}\IEEEauthorrefmark{3}}
\IEEEauthorblockA{\IEEEauthorrefmark{1}Chalmers | University of Gothenburg, Sweden}
\IEEEauthorblockA{\IEEEauthorrefmark{2}Radboud University Nijmegen, Netherlands}
\IEEEauthorblockA{\IEEEauthorrefmark{3}Ruhr University Bochum, Germany}
}






\newcommand{\vh}[1]{\rotatebox[origin=c]{90}{ \scriptsize #1}}
\newcommand{\f}[1]{{\tt #1}\xspace}
\newcommand\parhead[1]{\vspace{.26mm}\noindent\textbf{{#1}}}
\newcommand\parheadit[1]{\vspace{.26mm}\noindent\textit{{#1}}}
\newcommand{\Secref}[1]{Section\,\ref{#1}}
\newcommand{\Figref}[1]{Figure\,\ref{#1}}
\newcommand{\tabref}[1]{Table\,\ref{#1}}
\newcommand{\lstref}[1]{Listing\,\ref{#1}}
\newcommand{\appref}[1]{Appendix\,\ref{#1}}
\newcommand{\xcite}[1]{}

\definecolor{amethyst}{rgb}{0.6, 0.4, 0.8}
\definecolor{apricot}{rgb}{0.98, 0.81, 0.69}
\definecolor{applegreen}{rgb}{0.55, 0.71, 0.0}
\definecolor{antiquefuchsia}{rgb}{0.57, 0.36, 0.51}
\definecolor{amber}{rgb}{1.0, 0.49, 0.0}
\definecolor{atomictangerine}{rgb}{1.0, 0.6, 0.4}
\definecolor{ballblue}{rgb}{0.13, 0.67, 0.8}
\definecolor{babypink}{rgb}{0.96, 0.76, 0.76}
\definecolor{bittersweet}{rgb}{1.0, 0.44, 0.37}
\definecolor{codegreen}{rgb}{0,0.6,0}
\definecolor{codegray}{rgb}{0.5,0.5,0.5}
\definecolor{codepurple}{rgb}{0.58,0,0.82}
\definecolor{backcolour}{rgb}{0.95,0.95,0.92}
\definecolor{aureolin}{rgb}{0.99, 0.93, 0.0}
\definecolor{airforceblue}{rgb}{0.36, 0.54, 0.66}
\definecolor{aurometalsaurus}{rgb}{0.43, 0.5, 0.5}
\definecolor{bananayellow}{rgb}{1.0, 0.88, 0.21}
\definecolor{bazaar}{rgb}{0.6, 0.47, 0.48}
\definecolor{bostonuniversityred}{rgb}{0.8, 0.0, 0.0}
\definecolor{capri}{rgb}{0.0, 0.75, 1.0}

\newcommand{\marker}[1]{
    \begin{tikzpicture}
        \fill [#1,rounded corners=2, draw]
        (0,0) --
        ++(0.18,0) --
        ++(0,0.18) --
        ++(-0.18,0) --
        cycle
            {};
    \end{tikzpicture}
}
\newcommand{\expmkr}{\marker{capri}}
\newcommand{\pipemkr}{\marker{atomictangerine}}
\newcommand{\runmkr}{\marker{airforceblue}}
\newcommand{\modelmkr}{\marker{applegreen}}
\newcommand{\datamkr}{\marker{bananayellow}}
\newcommand{\teammkr}{\marker{bazaar}}
\newcommand{\hypmkr}{\marker{bostonuniversityred}}
\newcommand{\reportmkr}{\marker{amethyst}}

\newcommand{\ds}[1]{\textcolor{red}{\textbf{[DS]} #1}}

\lstdefinestyle{mystyle}{
    backgroundcolor=\color{backcolour},
    commentstyle=\color{codegreen},
    keywordstyle=\color{magenta},
    numberstyle=\tiny\color{codegray},
    stringstyle=\color{codepurple},
    basicstyle=\ttfamily\scriptsize, 
    breakatwhitespace=false,
    breaklines=true,
    captionpos=b,
    keepspaces=true,
    numbersep=5pt,
    showspaces=false,
    showstringspaces=false,
    showtabs=false,
    tabsize=2
}
\lstset{style=mystyle}

\hypersetup{
    colorlinks=true,
    linkcolor=black,
    filecolor=black,
    urlcolor=black,
}

\maketitle

\begin{abstract}
    \looseness=-1
Machine Learning (ML) techniques are becoming essential components of many software systems today, causing an increasing need to adapt traditional software engineering practices and tools to the development of ML-based software systems.
This need is especially pronounced due to the challenges associated with the large-scale development and deployment of ML systems. Among the most commonly reported challenges during the development, production, and operation of ML-based systems are experiment management, dependency management, monitoring, and logging of ML assets. In recent years, we have seen several efforts to address these challenges as witnessed by an increasing number of tools for tracking and managing ML experiments and their assets.
To facilitate research and practice on engineering intelligent systems, it is essential to understand the nature of the current tool support for managing ML assets. \textit{What kind of support is provided? What asset types are tracked? What operations are offered to users for managing those assets?
}
We discuss and position ML asset management as an important discipline that provides methods and tools for ML assets as structures and the ML development activities as their operations.
We present a feature-based survey of 17 tools with ML asset management support identified in a systematic search. We overview these tools' features for managing the different types of assets used for engineering ML-based systems and performing experiments.
We found that most of the asset management support depends on traditional version control systems, while only a few tools support an asset granularity level that differentiates between important ML assets, such as datasets and models.
\end{abstract}

\begin{IEEEkeywords}
    machine learning, SE4AI, asset management
\end{IEEEkeywords}

\section{Introduction}
\label{sec:introduction}
\noindent
\looseness=-1
An increasing number of software systems today implement AI capabilities by incorporating machine learning (ML) components. This growth increasingly demands using software-engineering methods and tools for systematically developing, deploying, and operating ML-based systems \cite{Kumeno2020, Arpteg2018, Polyzotis2017}. However, there are several difficulties associated with implementing these traditional methods in the ML application context. Arpteg et al. \cite{Arpteg2018} characterize these fundamental issues of engineering AI systems into development, production, and organizational challenges. They highlight experiment management, dependency management, monitoring, and logging as part of the core issues under the development and production challenges.
Moreover, engineers still face challenges to operationalize and standardize the ML development process \cite{Hill2016, Sridhar2018}. Hill et al.'s \cite{Hill2016} field interview study reveals that support for tracking and managing the different ML assets is essential, with engineers currently resorting to custom or ad hoc solutions, given the lack of suitable management techniques. All their interviewees also revealed they are limited to versioning only the source code and not other ML assets, while many specified that they adopted informal methods, such as emails, spreadsheets, and notes to track their ML experiments' assets.

\looseness=-1
ML practitioners and data scientists are tasked with managing ML assets when iterating over stages of the ML process lifecycle (described in \cref{subsec:ML-lifecycle}). The development iterations, which are usually repeated until the process results in an acceptable model, increase the number of generated models and their associated assets. Reports \cite{Vartak2016, Hill2016} show that ML practitioners usually generate hundreds of models before an acceptable model meets their target criteria. This iterative nature of developing ML systems contributes to the complexity of asset management in ML and calls for tool support to facilitate asset operations, such as tracking, versioning, and exploring. Specifically, we consider asset management in ML as a discipline that offers engineers the necessary management support for processes and operations on different types of ML assets. Various tools in the ML tools landscape, especially experiment management tools---the scope of our survey---offer asset management support.
\looseness=-1
Specifically, an ML experiment can be described as a collection of multiple iterations over stages of an ML process lifecycle towards a specific objective. The experiment management tools aim to simplify and facilitate the model-building and management processes by tracking essential ML assets. The assets commonly used in a model building process include the dataset and source code used in producing the model, the source code used for feature extraction on the dataset, the hyperparameters used during model training, and the model evaluation dataset. ML experiment management tools usually offer APIs via which users can log assets of interest and their relationships for operations such as versioning, exploration, and visualization. Some experiment management tools offer support for drawing insights from experiment outcomes through visualization, while some offer execution-related supports such as reproduction of ML experiments, parallel execution, and multi-stage executions.
Furthermore, note that ML practitioners and data scientists commonly use the term artifact to describe datasets and other model resources. In this work, we opt for the term \textit{asset} to describe all artifacts used (or reused) during the ML process lifecycle.

\looseness=-1
To facilitate research and practice on engineering ML-based systems, it is essential to understand the support that the current ML tool landscape offers to engineers and data scientists for managing the diversity of ML assets.
\textit{What support do they provide? What are the typical asset types they track? What operations are offered to engineer on the assets? What are their commonalities and variabilities?}

\looseness=-1
In this paper, we discuss and position asset management as an essential discipline to scale the engineering of ML experiments and ML-based systems. We survey asset management support in 17 contemporary experiment-management tools, identifying the types of assets supported, and the operations offered to engineers for managing ML assets. Specifically, we conduct a feature-based survey, which is essentially a domain analysis to identify the characteristics of an application domain (defined by the subject experiment-management tools we survey). We model these characteristics as features in a feature model \cite{kang.ea:1990:foda,nesic.ea:2019:fmprinciples}, an intuitive tree-like notation commonly used in software variability management\,\cite{apel.ea:2013:fospl,berger.ea:2020:emse}.
In the literature, such feature-based surveys have been performed before to compare the design space of technologies, such as model transformations\,\cite{transformationSurvey}, conversational AI systems\,\cite{aronsson2021maturity}, language workbenches\,\cite{erdweg2013languageworkbenches}, or variation control systems\,\cite{linsbauer2017gpce}. Our study contributes a feature-model-based representation of tools with support for ML asset management---particularly the ML experiment management tools---that captures the asset types and the supported operations. We address the following research questions:

\begin{itemize}
    \item \textbf{RQ1}: What asset types are tracked and managed in the subject tools?
    \item \textbf{RQ2}: What are the asset collection mechanisms used in collecting these assets?
    \item \textbf{RQ3}: How are the assets stored and version-controlled?
    \item \textbf{RQ4}: What are the management operations offered by the subject tools?
    \item \textbf{RQ5}: What integration support do they offer to other ML development systems?
\end{itemize}

\looseness=-1
Our study comprised collecting and selecting relevant tools for this study. We focused on experiment management tools, which typically offer management support for different ML asset types across stages of the ML process lifecycle. Specialized management tools, such as model-specific management tools (e.g., model registries and model databases \cite{Migliore2003}), dataset-specific management tools, pipeline or run orchestration-specific management tools, hyper-parameter management tools, visualization-specific tools, and experiment metadata databases \cite{Vanschoren2014} were beyond the scope of our study.

\looseness=-1
We hope that our study contributes to an increased empirical understanding of the solution space of ML asset management. ML practitioners and data scientists can use our survey results to understand the asset management features provided by contemporary  experiment management tools. Researchers can identify gaps in the tool support for ML asset management, as well as they can classify their new techniques against our taxonomy (the feature model). Lastly, we hope that our result will contribute towards building tools with improved ML management support that promote traditional software engineering methods in developing ML-based systems.


\section{Asset Management}
\label{sec:background}


\noindent
\looseness=-1
We now describe ML asset management as an essential discipline and position it by discussing background on ML process lifecycles and ML assets.

\subsection{ML Process Lifecycle}
\label{subsec:ML-lifecycle}
\begin{figure}[]
  \centering
  \includegraphics[width=\linewidth]{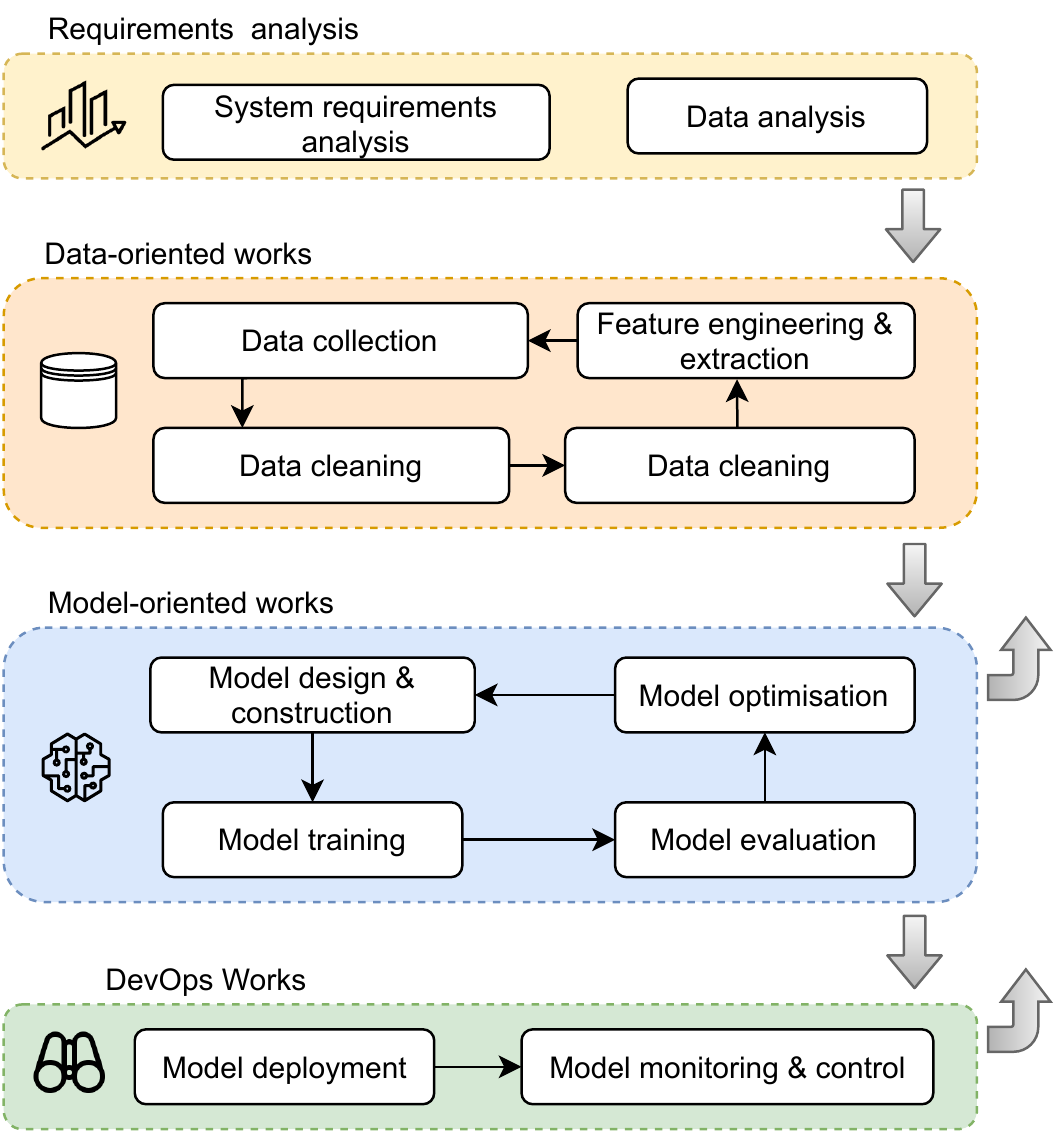}
  \caption{Stages in a typical ML workflow}
  \label{fig:supervised-ml-wf}
\end{figure}

\noindent
\looseness=-1
The traditional software engineering process \cite{Sarker2015} includes activities such as requirements analysis, planning, architecture design, coding, testing, deployment, and maintenance.
Similarly, ML follows a set of well-defined processes that are grounded in workflows designed in the data science and data mining context.
Examples include CRISP-DM \cite{Wirth2000}, KDD \cite{Fayyad1996}, and TDSP \cite{Microsoft2017}. \Cref{fig:supervised-ml-wf} shows a simplified workflow diagram of a supervised ML process lifecycle, structured along groups of development stages. The workflow consists of stages for requirements analysis, data-oriented works, model-oriented works, and DevOps works\,\cite{Kumeno2020}. The requirements analysis stages involve analyzing the system requirements and data, while the data-oriented stages include data collection, cleaning, labeling, and feature engineering or extraction. Model-oriented stages include model design, training, evaluation, and optimization. The DevOps stages include the deployment of ML models, monitoring and controlling of in-production models. \Cref{fig:supervised-ml-wf} illustrates multiple feedback loops (indicated by the upward arrows) from the model-oriented and DevOps stages to earlier stages. The feedback loop demonstrates iteration over sets of ML stages for a variable number of times until the process results in a model that meets a target objective. A need for asset management support is often attributed to the complexity and time overhead that arises with manually managing the large number of assets resulting from this iterative process \cite{Hill2016, Vartak2016, Schelter2018b}.

\subsection{ML Artifacts \& Assets}
\noindent
\looseness=-1
The term asset is conventionally used for an item that has been designed for use in multiple contexts \cite{iso_ieee_reuse_processes}, such as a design, a specification, source code, a piece of documentation, or a test suite. Consequently, we use the term asset for an artifact that, after its initial use, is retained for future use.
ML practitioners and data scientists often use the term artifact to describe required resources during an ML model development. These artifacts all qualify as assets in ML engineering because of the experimental nature and feedback loops in typical ML workflows, which requires keeping artifacts for future use.
Conventional software engineering often has fewer asset types to manage than the more extensive diversity of assets under ML engineering. Conventional software engineering mostly deals with textual artifacts, while ML includes additional artifact types, such as datasets, learned models, hyper-parameters, and model performance metrics\,\xcite{gollapudi2016practical}.

\subsection{Asset Management}
\label{subsec:management-tools}
\noindent
\looseness=-1
In this light, we define asset management as an essential discipline for scaling the engineering of ML-based systems and experiments:

\smallskip
\noindent\textbf{Definition 1} (Asset Management)\textbf{.} The discipline \textit{asset management} comprises methods and tools for managing \textit{ML assets} to facilitate activities involved in the development, deployment, and operation of ML-based systems. It offers \textit{structures} for storing and tracking ML assets of different types, as well as \textit{operations} that engineers can use to manage assets.
\smallskip

\looseness=-1
This definition emphasizes that establishing effective asset management requires efficient storage and tracking structures (e.g., data schemas, types, modular and composable units, and interfaces) as well as properly defined operations, which can be of different modalities (e.g., command-line tools or APIs allowing IDE integration). Asset management comprises activities pertaining to the practice areas
dataset management, model management, hyper-parameter management, process execution management, and report management.

\looseness=-1
\paragraph{Dataset management} The quality of datasets used in an ML model development plays a crucial role in the model's performance. Therefore, data understanding, preparation, and validation are crucial aspects of ML engineering. In this management area, tools (e.g., OrhpeusDB) focus on the ML lifecycle's data-oriented works and provide operations such as tracking, versioning, and provenance on dataset assets.

\looseness=-1
\paragraph{Model development management} Management tools in this area focus on model-oriented works of the ML lifecycle. They provide several supervised and unsupervised learning methods, such as classification, regression, and clustering algorithms to generate and evaluate ML models. The ML community has focused on model-oriented work, as witnessed by an extensive collection of available systems, frameworks, and libraries for model development (e.g., PyTorch, Scikit-Learn, or TensorFlow).

\looseness=-1
\paragraph{Model storage and serving management} Tools under this area focus on model-operation works of the ML process lifecycle. They provide efficient storage and retrieval of models to support the deployment, monitoring, and serving process. They provide information on the lineage of related assets and various evaluation performance of models (e.g., ModelDB).

\looseness=-1
\paragraph{Hyper-parameter optimization management} Searching or tuning for optimal hyper-parameters for a given ML task can be tedious. Tools in this area (e.g., Optuna) manage ML learning parameters and provide systematic ways to quicken the process of finding well-performing hyperparameters.

\looseness=-1
\paragraph{Pipeline \& run orchestration management} Tools in this area (e.g., AirFlow, Luigi, Argo) provides functionalities to orchestrate the automatic execution of ML lifecycle stages as described in \cref{fig:supervised-ml-wf}, from data collection to model serving. They often allow users to specify workflows as \textit{Direct Acyclic Graphs} (DAGs) to form collections of ML stages represented in a way that describes their dependencies with other ML assets. Also, they often adopt containerized technologies to support distributed and scalable ML operations. Training, testing, deploying, and serving models are examples of ML operations that benefit from using run orchestrators for faster model training and inference.

\looseness=-1
\paragraph{Reports \& visualizations} In this area, tools (e.g., TensorBoard, OmniBoard) present assets such as model evaluation metrics in graphical web dashboards to provide insight into ML experiments outcomes.

\section{Survey Methodology}
\label{sec:methodology}
\noindent
\looseness=-1
We now describe the methodology behind our survey of asset management capabilities within experiment management tools used in practice. Specifically, we systematically selected our candidate tools, and analyzed them to arrive at a feature model to answer our research questions.


\subsection{Tool Selection Process}
\noindent
Following established guidelines for systematic reviews\,\cite{Kitchenham2007}, we assessed existing ML experiment management tools for their capabilities. Selecting the tools was a four-step process:
\vspace{1mm}

\begin{itemize}
  \item First, we defined our search topic based on our research questions and accordingly chose our search terms.
  \item Second, we designed our search strategy and carried out our search on data sources (described shortly).
  \item Third, we identified and applied the initial selection criteria $C_1$ to collate a preliminary list of all identified ML management tools.
  \item Last, we identified and applied the additional selection criteria $C_2$ to arrive at the list of ML experiment management tools as the final candidate tools considered in this study.
\end{itemize}

\begin{figure}[b]
  \centering
  \includegraphics[width=\linewidth]{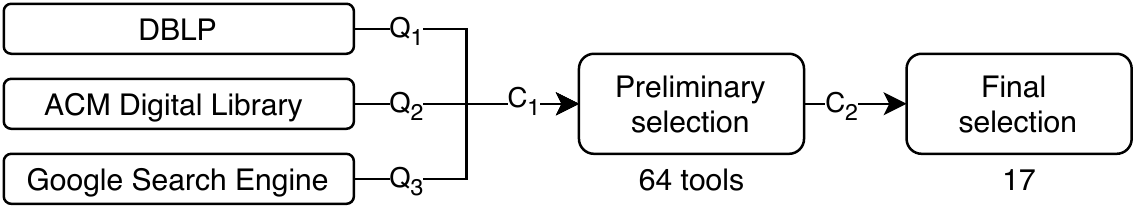}
  \caption{Overview of our selection process.}
  \label{fig:selection-process}
\end{figure}

\Cref{fig:selection-process} presents a summary of the selection process, and the following subsections provide further details on our selection process.

\subsubsection{Data Source and Search Strategy}
\looseness=-1
We employed three data sources in our work, namely DBLP, ACM digital library, and Google search engine.
Our search was conducted within the strategy guidelines presented by Kitchenham and Charters \cite{Kitchenham2007}. Accordingly, we excluded search terms such as ``asset'' from our search queries because we got several non-related results and instead used terms derived from our research questions and known related literature \cite{Isdahl2019, Ormenisan, Zaharia2018, Vartak2016, Ferenc2020, Weibgerber2019}. We also guided our search with our knowledge and experience of ML, and its application \cite{Idowu2014, Idowu2015, Idowu2016, struder2020feature, aronsson2021maturity, Abukwaik2018, queiroz2016prediction}. The following describes our search for each data source.

\looseness=-1
\textbf{$\textit{Source}_1$} (DBLP): First, we searched using DBLP,\xcite{dblp}—a comprehensive and high-quality bibliographic data source that only allows title search. From the research questions’ related terms, we created a search query, $Q_1$ ("machine learning" \& ("reproducible" | "reproducibility" | "reusable" | "provenance" | "lifecycle")), which resulted in a total of 46 publications.

\looseness=-1
\textbf{$\textit{Source}_2$} (ACM digital library): Since DBLP allows search on only the literature title, we searched both literature titles and abstracts using ACM digital library. Similar to the approach used for $Source_1$, we created a search query, $Q_2$ ("machine learning" AND ("reusable" OR "lifecycle" OR "reproducibility" OR "provenance")), based on relevant terms. Our search produced 12 and 127 results from the literature title and abstract search, respectively.

\looseness=-1
\textbf{$\textit{Source}_3$} (Google search engine):  Since we expected that many ML asset management tools used in practice are not documented in research publications, we performed an internet search using the search query, $Q_3$ ("machine learning") AND ("artifacts" OR "experiments") AND ("provenance" OR "versioning" OR "tracking" OR "history") AND ("Reusable" OR "reproducible") AND "management" AND ("framework" OR "tool" OR "platform"). From our search, we obtained 181 results. Note that there is a well-known phenomenon where Google search reports a significantly larger number of results than the actual result count. In our case, Google search initially reported over 2 million results, which later decreased to 181 results when we navigated to the last result page.


\looseness=-1
Using the literature we found from $\textit{Source}_1$ and $\textit{Source}_2$, we performed backward snowballing until we found no new relevant tools from the last paper. We manually filtered based on our selection criteria $C_1$ (described shortly) on results from all sources ($\textit{Source}_1$, $\textit{Source}_2$ and $\textit{Source}_3$) when collating asset management tools from the data sources. After that, we obtained a preliminary list of 66 ML asset management tools discarding duplicate entries. Finally, we selected using criteria $C_2$ (described shortly) to arrive at our final selection of tools with ML asset management support. 

\subsubsection{Selection Criteria}
\looseness=-1
As proposed by Kitchenham and Charters \cite{Kitchenham2007}, we describe the inclusion and exclusion criteria used in this survey to filter out and define the scope of tools that we analyzed. Our selection criteria ensured that we consider all relevant assets management tools to discover findings that pertain to our research questions. Since we filtered at two different stages of our selection process, we tagged the selection criterion $C_1$ and $C_2$. The prior indicates criteria applied when collating all tools with asset management support found from our data sources, while the latter indicates those applied to the preliminary selection to derive the experiment management tools (see \cref{fig:selection-process}).

\vspace{1mm}
\textbf{Inclusion criteria:} We considered the following:
\begin{itemize}
  \item Tools that covers any of the ML asset management areas described in \cref{subsec:management-tools}. ($C_1$)
  \item Tools with meaningful prominence measured by search trends and/or GitHub stars ratings. ($C_1$)
  \item  Tools with the primary purpose of ML experiment management (i.e., tools specifically designed to track and manage ML experiments and their assets) as we recognize them to be the most comprehensive with coverage of all asset management areas, and can provide insight into the ML asset management domain space when empirically examined. ($C_2$)
\end{itemize}

\vspace{1mm}
\textbf{Exclusion criteria:} We excluded the following:
\begin{itemize}
  \item Proposed frameworks or prototypes from literature. ($C_1$)
  \item Tools that lack well-defined documentation in English language. ($C_1$)
  \item General ML frameworks or ML development tools such as Scikit \cite{Pedregosa2011} and TensorFlow \cite{Abadi2016}. ($C_1$)
  \item Specialized tools for a single management area, such as dataset, ML model, hyper-parameter, pipeline, or execution orchestration management. ($C_2$)
\end{itemize}

\Cref{tab:selected-tools} shows our final 17 tools we evaluated in this work.


\aboverulesep=0ex
\belowrulesep=0ex

\begin{table}[]
    \caption{List of selected tools with ML asset management support.
    }
    \label{tab:selected-tools}
    \newcolumntype{m}{>{\hsize=.8\hsize}X}
    \newcolumntype{s}{>{\hsize=.06\hsize}X}
    \renewcommand{\arraystretch}{1.4}
    \rowcolors{2}{gray!15}{white}
    \begin{tabularx}{\columnwidth}{|X|X|}
        \rowcolor{gray!50}
        \toprule
        \textbf{Cloud Service }                                                             & \textbf{Software}                                                                                 \\
        \midrule

        Neptune.ml \scriptsize{(\href{https://neptune.ml}{netptune.ml})}                    & Datmo \scriptsize{(\href{https://github.com/datmo}{github/datmo})}                                \\
        Valohai \scriptsize{(\href{https://valohai.com}{valohai.com})}                      & Feature Forge \scriptsize{(\href{https://github.com/machinalis/featureforge}{github/machinalis})} \\
        Weights \& Biases \scriptsize{(\href{https://wandb.com}{wandb.com})}                & Guild \scriptsize{(\href{https://guild.ai}{guild.ai})}                                            \\
        Determine.ai \scriptsize{(\href{https://determined.ai}{determined.ai})}             & MLFlow \scriptsize{(\href{https://mlflow.org}{mlflow.org})}                                       \\
        Comet.ml \scriptsize{(\href{https://comet.ml/site}{comet.ml})}                      & Sacred \scriptsize{(\href{https://github.com/IDSIA/sacred}{github/IDSIA})}                        \\
        Deepkit \scriptsize{(\href{https://github.com/deepkit}{github/deepkit})}            & StudioML  \scriptsize{(\href{https://github.com/open-research}{github/open-research})}            \\
        Dot Science \scriptsize{(\href{https://dotscience.com}{dotscience.com})}            & Sumatra  \scriptsize{(\href{http://neuralensemble.org/sumatra/}{neuralensemble.org})}             \\
        PolyAxon \scriptsize{(\href{https://polyaxon.com}{polyaxon.com})}                   & DVC \scriptsize{(\href{https://dvc.org}{dvc.org})}                                                \\
        Allegro Trains \scriptsize{(\href{https://github.com/allegroai}{github/allegroai})} & -                                                                                                 \\
        \bottomrule
    \end{tabularx}

    \vspace{5pt}
\end{table}


\subsection{Analysis of Identified Tools}
\noindent
\looseness=-1
This study aims to recognize the characteristics that differentiate our subject tools using features\,\cite{berger.ea:2015:feature} and to represent them in a feature model\,\cite{kang.ea:1990:foda,nesic.ea:2019:fmprinciples}. This analysis process has been divided into different stages to study our candidate tools’ management capabilities and build a resulting feature model. Our analysis is based on information found in publicly available documentation for each of the tools, and in a few cases, we had to test the tools for their available functionalities when needed practically.

\par
\looseness=-1
First, we have performed an initial analysis of a single ML tool to identify its supported ML asset types, the collection approaches, the storage options, supported asset operations, and integration capabilities.
We partly established the terminologies to be used in our models. This stage produced our baseline version of the feature model that we present as our contribution.

\par
\looseness=-1
Second, building on the first draft version, we adopted an iterative process to evaluate additional tools while modifying terminologies and the model structure to accommodate variations from the new tools being assessed.

\par
\looseness=-1
Lastly, at the end of the final iteration, all the authors met to review the latest structure for direct feedback on terminologies used and the feature model. We integrated this feedback into our survey to arrive at the contribution of this study.


\section{Asset Management Features}
\label{sec:feature-model}
\begin{figure}[tb]
  \centering
  \includegraphics[trim={0, 160, 0, 60}, width=\linewidth]{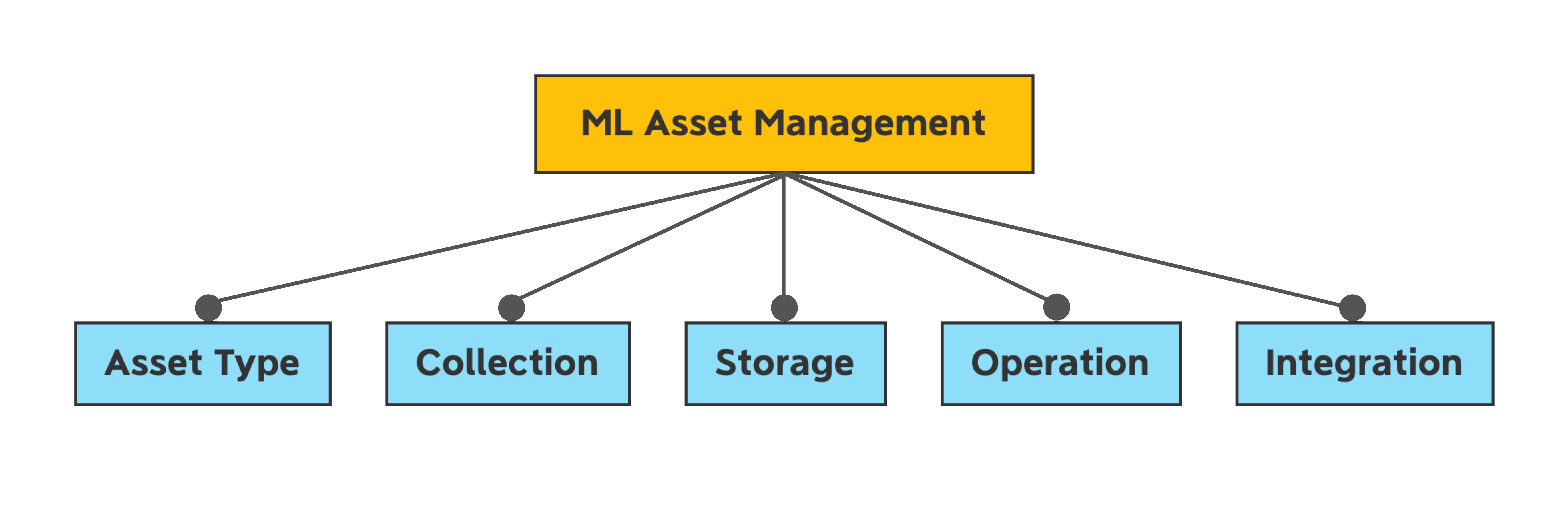}
  \smash{\begin{minipage}{7.5cm}\fontsize{6}{6}
    \leftline{$\multimapdotinv \textsf{Represented in}$}
    \leftline{$~~\textsf{all subjects}$}
	  \vspace{3.5cm}
  \end{minipage}}
  \caption{Main ML asset management features.}
  \label{fig:mlassets-top-level}
\end{figure}

\noindent
\looseness=-1
We propose a feature model—outlined in Figs.~\ref{fig:mlassets-top-level}-\ref{fig:integration-model}—to characterize and describe the asset management support of our subjects.
The top-level features—\f{Asset Type}, \f{Collection}, \f{Storage}, \f{Operation} and \f{Integration}—capture the core functionalities of the subjects in our study. \f{Asset type} outlines the data types that are tracked by our subjects; \f{Collection} describes how the assets are collected; \f{Storage} describes how the assets are stored and versioned; \f{Operation} specifies what operation types are supported; and \f{Integration} shows the subjects' integration support to other systems.

We describe these top-level features and their corresponding sub-features in the following subsections.

\subsection{Asset Type (RQ1)}
\label{subsec:AssetType}
\begin{figure}[b]
  \centering
  \includegraphics[trim={0, 105, 0, 110}, clip, width=.8\linewidth]{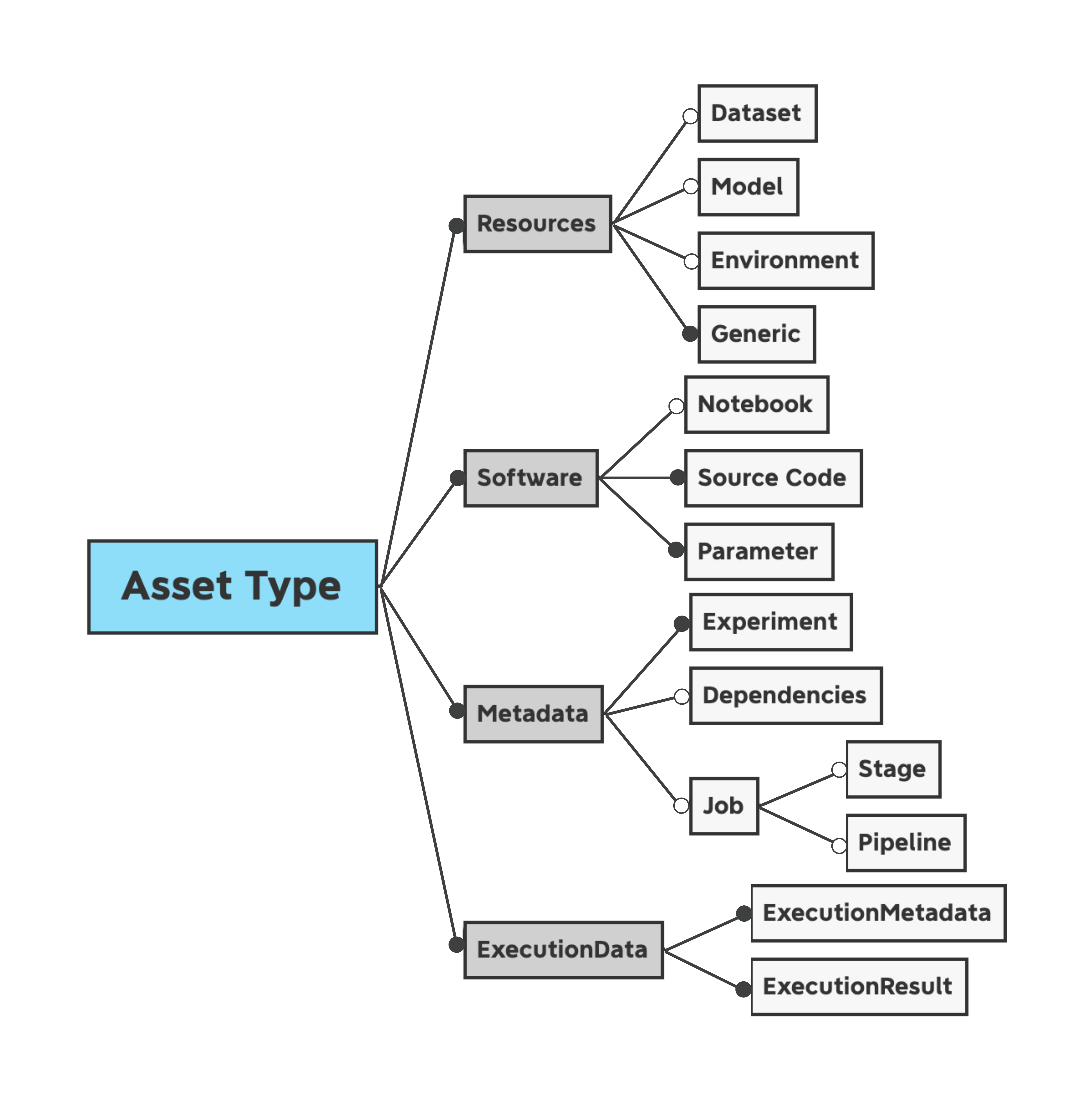}
  \smash{\begin{minipage}{7.5cm}\fontsize{6}{6}
	  \leftline{$\multimapdotinv \textsf{Represented in all subjects}$}
	  \leftline{$\multimapinv \textsf{Represented in some subjects}$}
	  \vspace{12.2cm}
  \end{minipage}}
  \caption{Asset type feature model: A representation of the data types tracked by the subjects under study.}
  \label{fig:asset-model}
\end{figure}

\noindent
\looseness=-1
Following the definition of the term asset, as presented in \cref{sec:background}, we define the feature \f{Assets Type} as the set of data types tracked and managed by our subjects. Contrary to traditional software engineering, whose primary asset type is source code, ML engineering has more diversified asset types, such as datasets, hyper-parameters used in training, trained models, and evaluation metrics. As shown in \cref{fig:asset-model}, \f{Resources}, \f{Software}, \f{Metadata}, and \f{ExecutionData} are sub-features of \f{Asset type}.

\looseness=-1
\subsubsection{Resources} Resources, also commonly referred to as 'artifacts' by many of the subjects, are the asset types required as input or produced as output from an ML workflow’s stage (see \cref{fig:supervised-ml-wf}). The subjects track resources with varying abstraction levels from specific asset types to \f{Generic} ones. We identified \f{Dataset} and \f{Model} as the most critical resource types. Many of them allow users to log the location and hash of data stored on local filesystems or cloud storage systems, such as AWS S3, Azure Storage, and Google Cloud Storage.

\looseness=-1
\paragraph{Dataset} The feature \f{Dataset} is available for subjects that identify datasets as an asset type. Data is an essential asset type in machine learning. Most of the ML workflow's stages, such as data collection, data transformation, feature extraction, model training, and evaluation, are data-dependent. The version of datasets used in ML workflow stages can be tracked to provide data lineage information for ML experiments.

\looseness=-1
\paragraph{Model} The feature \f{Model}  is available for subjects that identify models as an asset type. ML models are created by learning from datasets using learning algorithms provided by ML development frameworks. Models are tracked along with their associated assets to facilitate result analysis, such as comparing models from different experiment runs.

\looseness=-1
\paragraph{Environment} Tatman et al. \cite{Tatman2018a} reveal that sharing an environment with source code and dataset provides the highest level of reproducibility. With feature \f{Environment}, users can track environment resources such as Docker containers or Conda environments as experiments’ assets to ensure reproducible ML experiments.

\looseness=-1
\paragraph{Generic} Several subjects lack dedicated support for tracking \f{Dataset} and \f{Model} types; instead, they provide a “one-size-fits-all” tracking of generic resources. Consequently, subjects with feature \f{Generic} can track all asset types that are required or generated during an ML experiment without differentiating them.

\looseness=-1
\subsubsection{Software} This represents the software implementation of the ML process. \f{Software} typically involves the implementation of one or more stages of an ML workflow, and heavily relies on the supporting ML frameworks or model development tools that provide a collection of general ML techniques such as SciKit-Learn,\xcite{\url{https://scikit-learn.org/stable/}} PyTorch,\xcite{\url{https://pytorch.org}} TensorFlow,\xcite{\url{https://www.tensorflow.org}} and Keras\xcite{\url{https://keras.io}} .
We identified the sub-features of \f{Software} as \f{Notebook}, \f{SourceCode}, and \f{Parameter}.

\looseness=-1
\paragraph{Notebook} Similar to source code, notebooks contain the implementation to carry out specific ML operations. \f{Notebooks}, written in multiple execution cells, are usually used for small-scale, exploratory, and experimental ML tasks, where it is difficult to achieve acceptable software engineering practices such as modular design or code reuse. Notebooks (e.g., Jupyter \cite{Kluyver2016}) are crucial for reproducible machine learning workflows that require literate and interactive programming. The \f{Notebook} feature indicates the support to track notebooks as an asset type. Users can track or version notebooks using snapshots or through Notebook checkpoints.

\looseness=-1
\paragraph{Source Code} This feature represents text-based files with implementation to carry out specific ML operations. Managing source code (or scripts) is generally less challenging than notebook formats for functional and large-scale engineering of ML-based systems because of available IDEs to support code assistance, dependency management and debugging \cite{chattopadhyay2020}. Besides, source code are text-based files; therefore, they are easily version-controlled using traditional repositories. Consequently, ML practitioners and data scientists working on large scale systems often employ notebooks for initial ML experimentation and later convert them to source code files.

\looseness=-1
\paragraph{Parameter} Hyper-parameters are parameters utilized to control the learning process of an ML algorithm during the training phase of a model from a dataset (e.g., learning rate, regularization, and tree depth). Hyper-parameters are commonly tracked to facilitate the analysis of ML experiments' results. Some subjects (e.g., Comet.ml, Polyaxon, and Valoh.ai) provide hyper-parameter tuning and search features to facilitate the model-oriented stages of an ML workflow. In addition to hyper-parameters, the \f{Parameter} asset type also represents other configurable parameters that may influence an ML process.

\looseness=-1
\subsubsection{Metadata} Metadata is a vital part of information management systems since it allows the semantic description of entities. In our context, as a sub-feature of \f{Asset Type}, \f{Metadata}  represents the descriptive and structural static information about ML experiments, their dependencies, and how they are executed.

\looseness=-1
\paragraph{Experiment} The \f{Experiment} represents the main asset-type to which other assets are associated. It is the core abstraction of experiment management tools. Other tracked assets are usually linked with an \f{experiment}.

\looseness=-1
\paragraph{Dependencies} These are metadata information about the environment dependencies of an experiment. Examples include environment variables, host OS information; hardware details; Python libraries, and their versions.

\looseness=-1
\paragraph{Jobs} This feature represents the execution instructions of an ML experiment and how assets defined by \f{Resources} and \f{Software} should be used during execution. There is usually a 'one-to-one' or 'one-to-many' relationship between an \f{Experiment} and their \f{Jobs}. Following from the ML workflow, the feature \f{Job} can be described as a \f{Stage} or a \f{Pipeline}.  \cref{lst:dvc-example} shows the representation of a stage and pipeline in DVC. Using CLI command as execution instruction by some subjects can also be seen as a different form of \f{Job} representation.

\begin{lstlisting}[label=lst:dvc-example, float, language=Python, caption=An example of $\f{Metadata}$ representation of a $\f{Pipeline}$ with two $\f{Stages}$ in our subject DVC.]
stages: # Pipeline
  prepare: # Stage
    cmd: python src/prepare.py data/data.xml
    deps:
    - data/data.xml
    - src/prepare.py
    params: # Configuration
    - prepare.seed
    outs:
    - data/prepared
  featurize: # Stage
    cmd: python src/featurization.py data/prepared data/features
    deps:
    - data/prepared
    - src/featurization.py
    params: # Configuration
    - featurize.max_features
    - featurize.ngrams
    outs:
    - data/features
\end{lstlisting}

\begin{itemize}[leftmargin=*]
  \item[--] A \f{Stage} is a basic reusable phase of ML workflow as illustrated in \cref{fig:supervised-ml-wf}. They are defined with pointers to its required assets, such as \f{SourceCode}, \f{Parameters}, and input \f{Resources} (e.g., datasets).

  \item[--] A \f{Pipeline} represents a reusable relationship between multiple stages to produce an ML workflow variants described in \cref{fig:supervised-ml-wf}. ML pipelines are usually built as dependency graphs, where it uses input and output resources as dependencies between stages. In \cref{lst:dvc-example}, a dependency graph is represented with \textit{featurize} stage which depends on the output of the \textit{prepare} stage.
\end{itemize}

\looseness=-1
\subsubsection{ExecutionData} This feature represents execution-related data that are tracked explicitly or automatically during the execution of an ML experiment. We identified \f{ExecutionMetadata} and \f{ExecutionResult} as sub-feature of \f{ExecutionData}.

\looseness=-1
\paragraph{ExecutionMetadata} This feature represents information about the execution process that is usually captured while execution (e.g., model training) is ongoing. These include terminal outputs, execution duration, events, statuses, and hardware consumption, such as CPU, GPU, and memory utilization.

\looseness=-1
\paragraph{ExecutionResult} This feature represents the assets generated as the results of an experiment. For a model training stage, this usually refers to evaluation metrics and can be tracked in different forms based on the ML task (e.g., sensitivity or ROC values for classification tasks; MSE, MAPE, or $R^2$ for regression tasks). For model training and data-oriented stages, \f{Model} and \f{Dataset} assets types are the results; hence, this indicates a relationship between the feature \f{ExecutionResult} and feature \f{Resources}.

\looseness=-1
\summary{What asset types are tracked and managed? (RQ1)}{
Our subjects support resources (including datasets, models, environments, and generic resources); software (including notebooks, source code, and parameters); metadata (including experiment, dependencies, stages, and pipelines); and execution data (including execution metadata and results).
}

\subsection{Collection (RQ2)}
\label{subsec:Collection}
\begin{figure}[tb]
  \centering
  \includegraphics[trim={0, 120, 0, 120}, clip, width=0.9\linewidth]{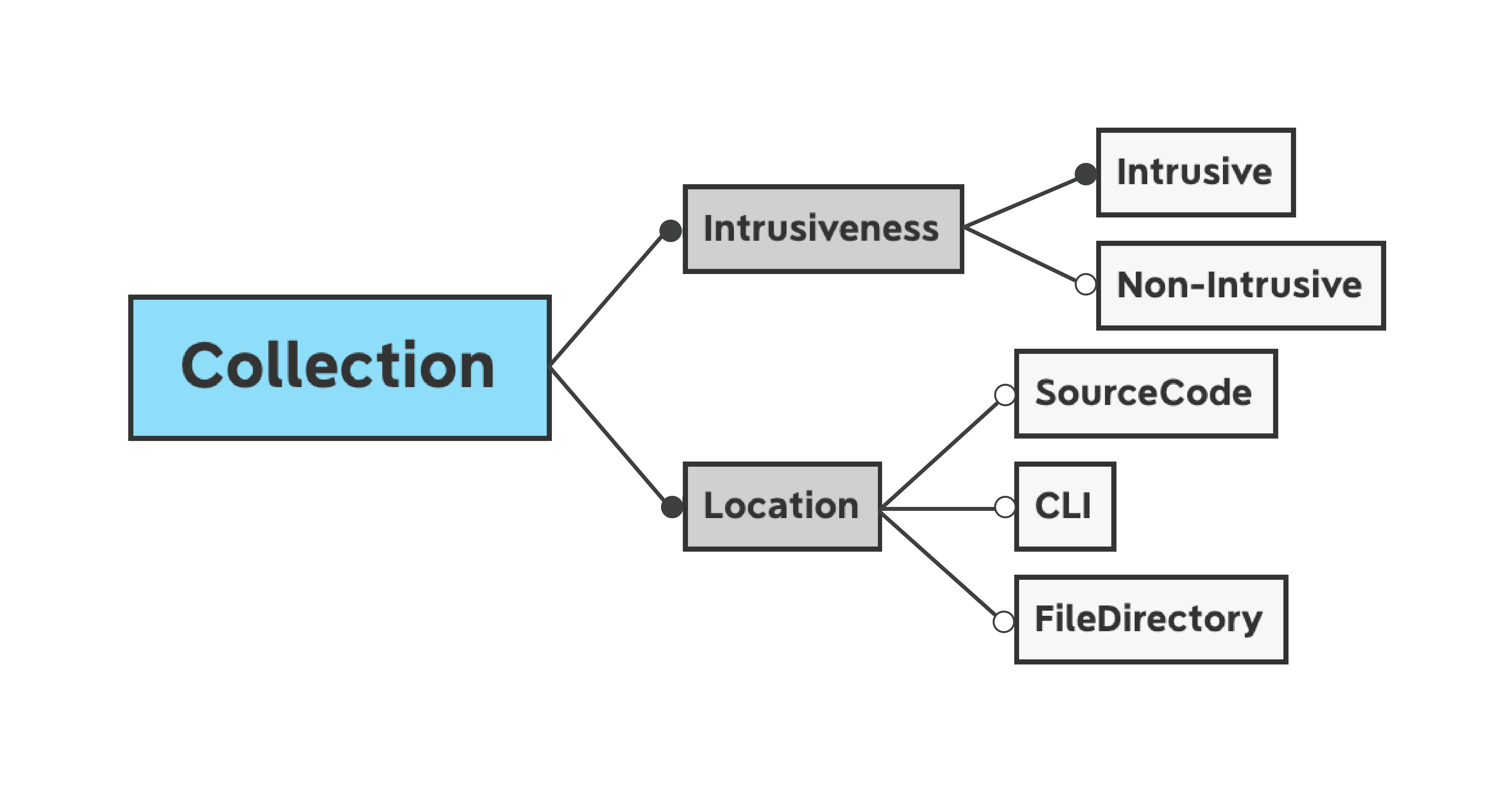}
  \caption{Collection feature model: A representation of collection features used in tracking the asset types described in \cref{subsec:AssetType}.}
  \smash{\begin{minipage}{7.5cm}\fontsize{6}{6}
	  \leftline{$\multimapdotinv \textsf{Represented in all subjects}$}
	  \leftline{$\multimapinv \textsf{Represented in some subjects}$}
	  \vspace{7cm}
  \end{minipage}}
  \label{fig:collection-model}
\end{figure}

\noindent
\looseness=-1
Feature \f{Collection}, shown in \cref{fig:collection-model}, represents the options provided by our subjects to track ML assets. The feature \f{Intrusiveness} shows the level of the explicit declaration required to collect the assets;  while feature \f{Location} represents the point where assets are being collected.

\looseness=-1
\paragraph{Intrusiveness} This describes the amount of instrumentation required by users to track assets. The \f{Intrusive} collection is invasive and requires users to add specific instructions and API calls in source code to track or log desired assets. In contrast, the \f{Non-Intrusive} collection automatically tracks or logs assets without the need for explicit instructions or API calls.

\looseness=-1
\paragraph{Location} This describes the collection point of assets. Assets can be extracted from \f{SourceCode}, \f{CLI} or \f{FileDirectory}. The common collection point across the considered subjects is the \f{SourceCode}, where the subjects provide a library and API that can be invoked to log desired assets within source code implementation. For collection at the \f{CLI}, subjects that are invoked via CLI commands allow users to specify pointers to assets as command arguments. In some subjects, CLI output are parsed to obtain the \f{ExecutionData's} assets. With the \f{FileDirectory} approach, subjects monitor assets from structured or instrumented file systems. Assets collected through this method are usually \f{Non-Intrusive} as modifications are automatically tracked.

\looseness=-1
\summary{How are assets collected? (RQ2)}{%
The collection approach can be intrusive or non-intrusive. Assets are commonly collected from source code; other collection locations include CLI arguments or logs, and instrumented file systems.
}

\subsection{Storage (RQ3)}
\label{subsec:AssetType}
\begin{figure}[]
  \centering
  \includegraphics[trim={0, 65, 0, 65}, clip, width=0.7\linewidth]{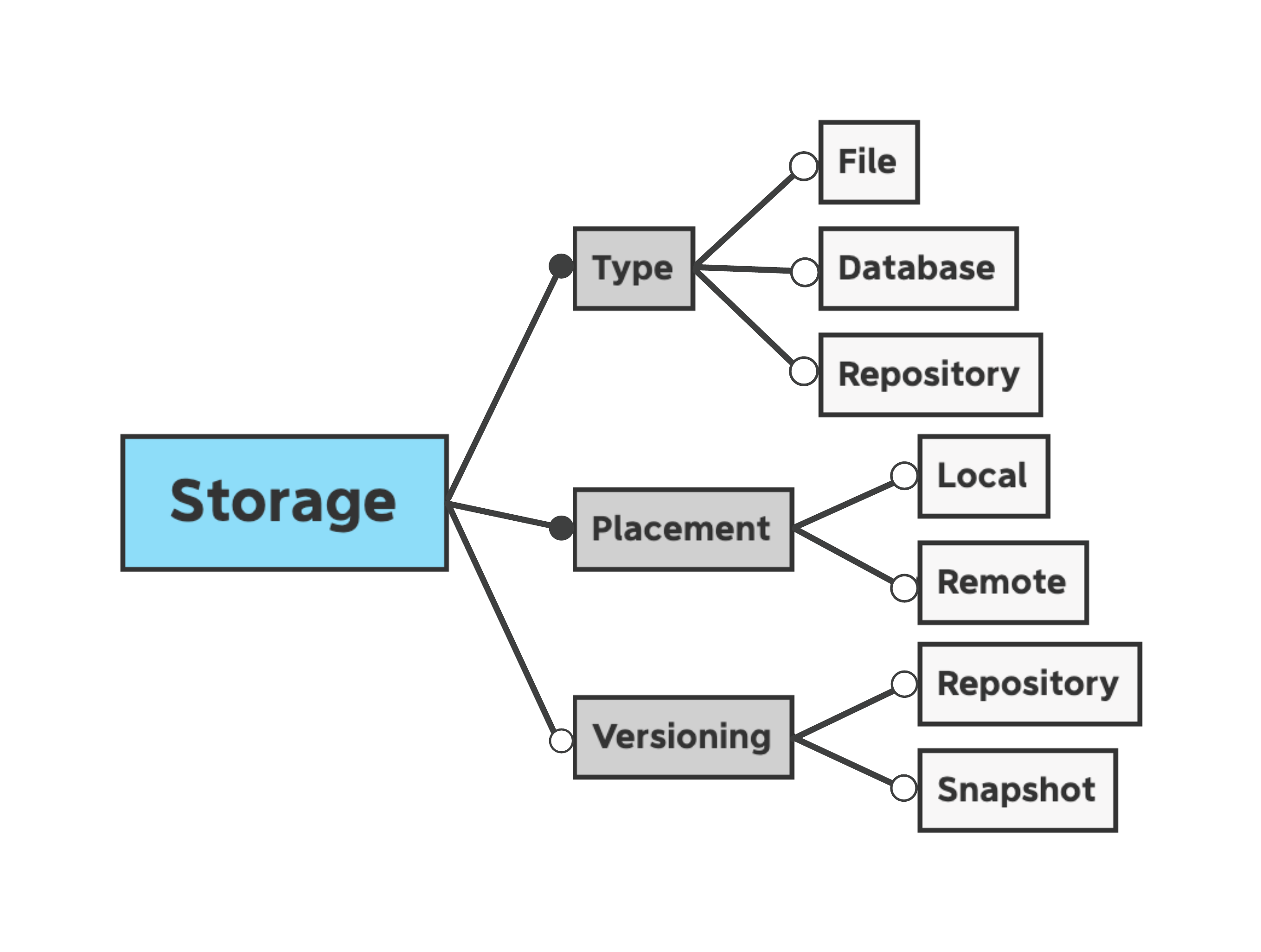}
  \caption{Storage feature model: A representation of the storage feature observed from the subjects under study.}
  \smash{\begin{minipage}{7cm}\fontsize{6}{6}
	  \leftline{$\multimapdotinv \textsf{Represented in all subjects}$}
	  \leftline{$\multimapinv \textsf{Represented in some subjects}$}
	  \vspace{8cm}
  \end{minipage}}
  \label{fig:storage-model}
\end{figure}

\noindent
The feature \f{Storage} describes how the assets are stored and version controlled, and \cref{fig:storage-model}  shows its sub-features.

\looseness=-1
\paragraph{Storage Type} The feature \f{Storage} is fundamental to the observed subjects, especially the cloud services, which also provides cloud storage capabilities for generated assets. We identified \f{File}, \f{Database}, and \f{Repository}-based type of storage.

\looseness=-1
\paragraph{Storage Placement} The primary option provided for storage placement are \f{Local} and \f{Remote} placement with respect to the management tools. Assets stored remotely are usually tracked through identifier pointers and are transferred or fetched for processing on demand. This option is suitable for large files and scenarios where users require easy access from cloud-hosted services, such as notebooks and cloud computing infrastructure.

\looseness=-1
\paragraph{Versioning} Support for \f{Versioning} is required to track the evolution of assets by keeping versioned assets during ML development. This feature is supported either by delegating asset versioning to traditional version control \f{Repository}, or via \f{Snapshot} of assets at each checkpoint. For the \f{Snapshot} approach, subjects independently create and track versions of assets, such as \f{Notebooks} and sub-features of \f{Resources}. In contrast, the feature \f{Repository} represents the collection and tracking of repository information (e.g., the commit hash and commit messages) with the associated experiment.

\looseness=-1
\summary{How are assets collected? (RQ3)}{%
The assets are either stored in file systems, databases, or repositories, either locally or remotely, while assets are version-controlled internally through snapshots or delegated to existing traditional repositories, such as Git.
}

\subsection{Operations (RQ4)}
\label{subsec:Operation}
\begin{figure}[tb]
  \centering
  \includegraphics[trim={0, 50, 0, 50}, clip, width=0.8\linewidth]{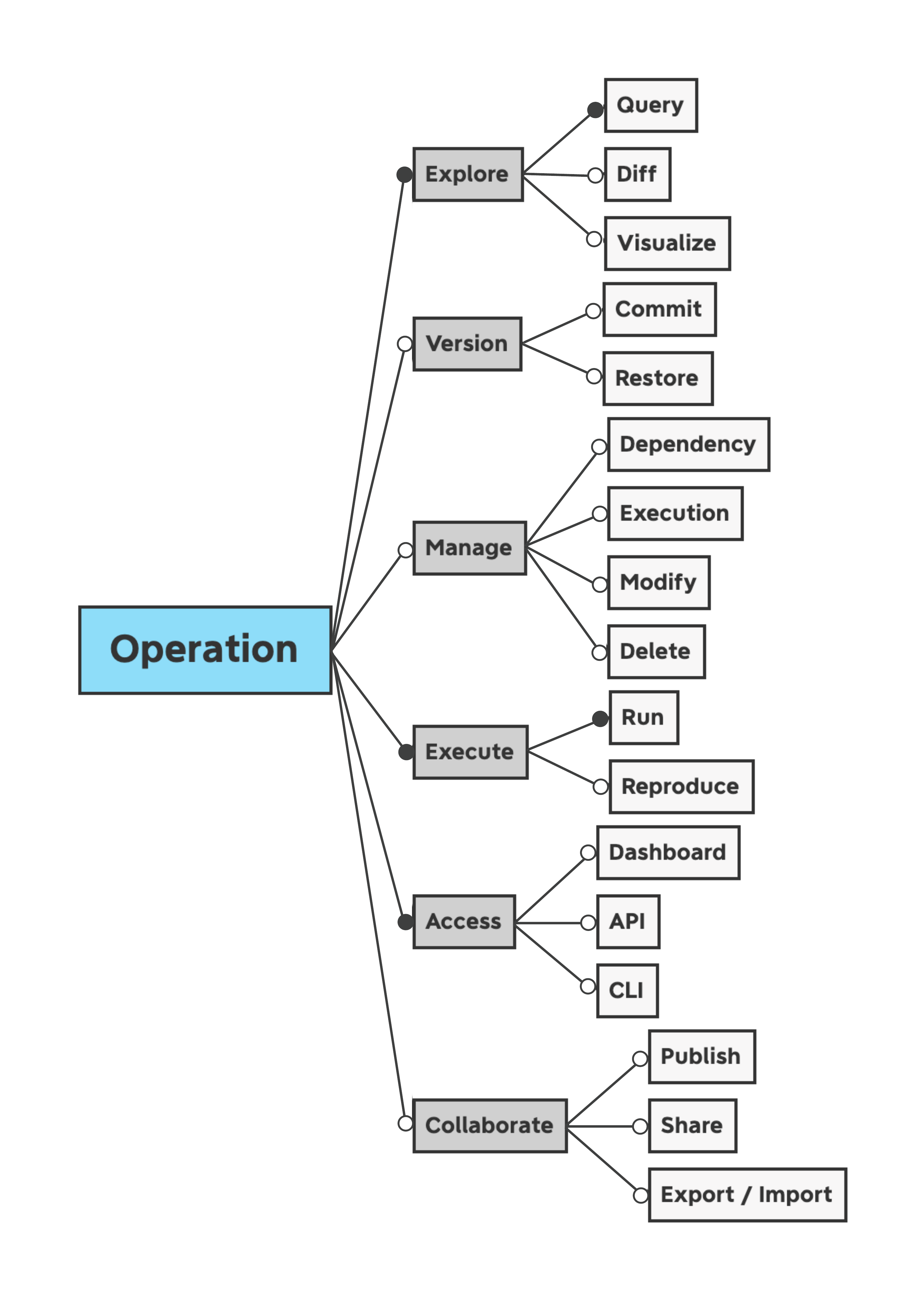}
  \caption{Operation feature model: A representation of operations offered by the subjects under study.}
  \smash{\begin{minipage}{7cm}\fontsize{6}{6}
	  \leftline{$\multimapdotinv \textsf{Represented in all subjects}$}
	  \leftline{$\multimapinv \textsf{Represented in some subjects}$}
	  \vspace{18cm}
  \end{minipage}}
  \label{fig:operation-model}
\end{figure}

\noindent
\looseness=-1
We identified several operations supported by our subjects and represent them by the feature \f{Operation}. The \f{Execute} and \f{Explore} operations are the primary features supported by all subjects. \cref{fig:operation-model} shows the sub-features of \f{Operation}.

\looseness=-1
\paragraph{Explore} The feature \f{Explore} represents the presence of operations that help derive insight and analyze experiment results. The subjects support various ways to \f{Query} assets, from listing all experiment assets to selection based on model performance filters. \f{Diff} indicates the presence of diffing between two or more assets while \f{Visualize} indicates the use of graphical presentations (e.g., charts) of experiments and their associated assets.

\looseness=-1
\paragraph{Version} This feature represents versioning-related operations. The feature \f{Commit} represents the presence of operations to create experiment checkpoints; while \f{Restore} indicates an operation to revert to an earlier version.

\looseness=-1
\paragraph{Manage} The feature \f{Dependency} represents the presence of dependency management, which is often supported by tools with \f{Stage} and \f{Pipeline} features as multi-stage execution. As an example, the \textit{deps} in \cref{lst:dvc-example} indicate the assets that are required for each stage. The \f{Execute} feature indicates the management of how ML experiments are being executed, similar to run-orchestration-specific tools. Other sub-features of the feature \f{Manage} include \f{Modify} which provides the option to revise already logged assets (mostly the metadata of experiments), while  \f{Delete} offers the option to remove already stored assets of an experiment.

\looseness=-1
\paragraph{Execute} The feature execute represents operations that invokes ML experiment via a defined entry-point. The features \f{Run} and \f{Reproduce} allow the execution of new and the reproduction of prior experiments, respectively.

\looseness=-1
\paragraph{Access} The mostly supported approach to \f{Access} stored assets is via graphical \f{Dashboards}. Other means of access include \f{API}, which provides REST interfaces or programming language APIs to access assets stored by the subject; the feature \f{CLI} exists for subjects that provide CLI commands for asset management.

\looseness=-1
\paragraph{Collaborate} This feature represents the presence of collaboration features, which are targeted at teams that need to share assets and results among the team members. User can \f{Publish}, \f{Share}, \f{Export} or \f{Import} experiment results or other required assets.

\looseness=-1
\summary{What are the supported operations? (RQ4)}{%
The supported operations allow users to explore assets using queries, comparison and, visualization of assets for insights. Other operations allow users to version assets, manage dependencies of assets and how they are executed. Users can also modify assets; run or reproduce experiment, access stored assets, and collaborate by publishing, sharing, exporting, or importing of assets.
}

\subsection{Integration (RQ5)}
\label{subsec:Integration}
\begin{figure}[tb]
  \centering
  \includegraphics[trim={0, 125, 0, 125}, clip, width=0.7\linewidth]{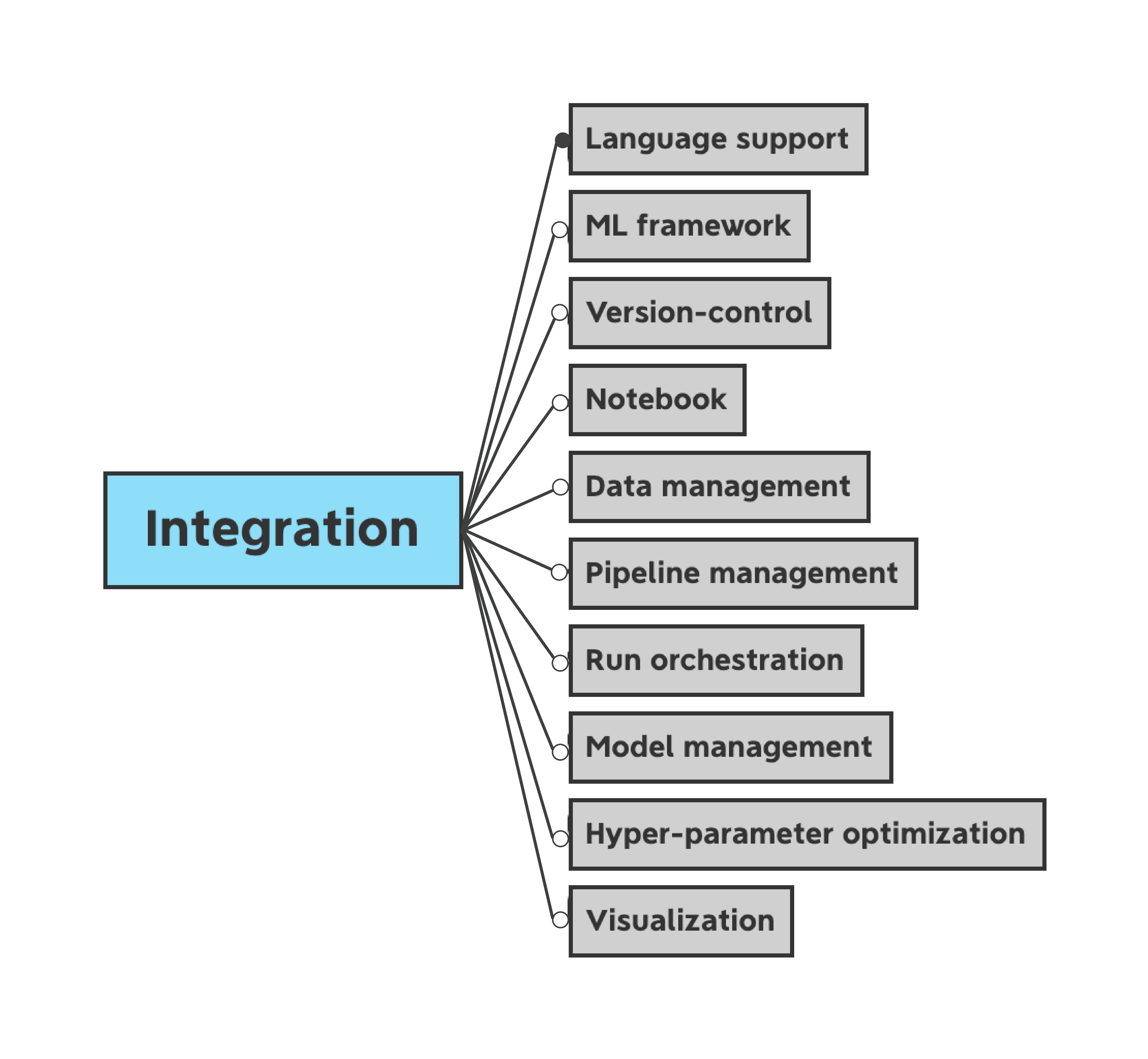}
  \caption{Integration feature model: A representation of the integration support offered by the subjects under study.}
  \smash{\begin{minipage}{7cm}\fontsize{6}{6}
	  \leftline{$\multimapdotinv \textsf{Represented in all subjects}$}
	  \leftline{$\multimapinv \textsf{Represented in some subjects}$}
	  \vspace{10cm}
  \end{minipage}}
  \label{fig:integration-model}
\end{figure}

\noindent
\looseness=-1
The experiment management tools are usually a piece of a larger toolchain. In large-scale ML production, several ML tools and frameworks are employed to develop, deploy, operate, and monitor ML models.
The \f{Integration} feature, illustrated in \cref{fig:integration-model}, identifies the integration support type commonly provided by our subjects.

\looseness=-1
The feature \f{Language support} indicates support for programming languages, most notably Python. Many of our subjects support asset \f{collection} from source code; hence, they provide support for common \f{ML Frameworks} such as TensorFlow and SciKit-Learn. Such support allows some of our subjects to offer non-intrusive asset collection. Subjects that delegate to or depend on traditional versioning systems to version assets provide \f{Version-control} integration support. Certain subjects also provide integration for \f{Notebooks} in the form of plugins. \f{Data Management} indicates the integration support to retrieve and store assets from and to data specific tools and frameworks. The integration supports for workflow and execution orchestration systems are indicated by \f{Pipeline Management} and \f{Run Orchestration}; while the integration support for model-specific management tools is indicated by feature \f{Model Management}. \f{Hyper-parameter Optimization} indicates the presence of support for parameter search and tunning tools; while \f{Visualization} indicates the presence of integration for visualization libraries, such as Omniboard,\xcite{\url{https://github.com/vivekratnavel/omniboard}} and TensorBoard\xcite{\url{https://www.tensorflow.org/tensorboard}}.

\looseness=-1
\summary{What integration support is offered? (RQ5)}{%
The subjects provide APIs, most frequently, Python-based ones. Further support includes ML frameworks, versioning systems, notebooks, data and model management tools, pipeline management and run orchestration tools, hyper-parameter optimization tools, and visualization libraries.
}

\section{Discussion}
\label{sec:discussion}

\noindent
\looseness=-1
\parhead{Version Control Systems.}
\looseness=-1
Most of the subjects in this study delegate the versioning of ML assets to version control systems, such as Git. This approach increases tooling complexity for users and can be a deterring factor for adoption. As shown in existing studies \cite{Hill2016, Vartak2016}, a large percentage of data scientists and ML practitioners still use solutions best described as ad hoc, while failing to achieve systematic ways of managing ML assets. A homogeneous way of ML asset management, where a user can manage all asset types from a single interface, will be a step towards encouraging the adoption of asset management tools. To achieve such, it would seem reasonable to extend traditional version control systems to a system that can support more ML asset types beyond text-based ones. Besides, the willingness to adopt these tools can be affected by the development context. For example, practitioners working on large ML projects that require collaboration and often generate hundreds of models are more likely to utilize asset management systems.

\parhead{Implicit Collection.}
\looseness=-1
We observe that most subjects' asset collection methods are intrusive, i.e., it requires the users to instrument their source code to track asset information. This method is tedious and error-prone and can also be a deterring factor in adopting tools with management support. The non-intrusive asset collection methods---supported by subjects such as MLFlow\xcite{mlflowweb} and Weights \& Biases\xcite{wand} for specific ML general frameworks, such as TensorFlow and SciKit Learn---eliminate these drawbacks. Ormenisan et al. \cite{Ormenisan} also proposed an implicit method of asset collection based on an instrumented file system and promises to solve issues associated with the intrusive asset collection methods.

\parhead{Reusability.}
\looseness=-1
Reproducibility is one of the main objectives of using ML experiment management tools, and there is a significant presence of such features across most subjects in this study. In contrast, the reusability, which can significantly impact model development, is not often addressed. Few of the subjects provide a limited operation to reuse assets under the \f{Manage::Execution} operation, where users can define how experiment jobs should be executed. We observe that only the asset type \f{Job::Stage} and \f{Job::Pipeline} are provided with such operation, where the execution path of an ML pipeline skips unmodified stages when reproducing a pipeline. Reusability of more ML assets can significantly reduce model development time, enhance asynchronous collaboration in development teams, and motivate ML model evolution use-cases. We expect to see more tools addressing the reusability challenge of ML assets in the future.







\section{Threats to Validity}
\label{sec:threat-to-validity}
\noindent

\parhead{External Validity.}
\looseness=-1
The majority of tools surveyed is Python-based, and we identify this as a threat to external validity, since it may impact result generalization to other tools. However, we believe our feature model is valid for most ML experiment management cases, since Python will remain the most widely used language in ML development for the foreseeable future, among others, for its abundance of available ML-related packages. The chosen terminologies of the tools we observed vary based on the tools' target groups (e.g., ML practitioners, data scientists, researchers) or experiment type (e.g., multi-purpose, machine learning, or deep learning experiment). To enhance external validity, we adopted broad terminologies through multiple iterations of analysis per tool to ensure uniformity and generalization across all subjects.

\parhead{Internal Validity.}
\looseness=-1
We manually selected the considered tools. One threat to the internal validity might be that the collection and filtering are subjective to individual opinion. We minimize this threat by validating our selection with information from the grey literature, such as market analysis reports.
Since we consider a rapidly evolving technology landscape,
we provide the snapshot date of accessed information.
Furthermore, our internet exploration using the Google search engine is prone to varying results based on user, time, and search location---personalized user experience. This issue threatens the ability to reproduce the same search by other researchers. To mitigate these threats, we relied on multiple data sources to increase the reliability of our data collection process. For the cloud-based services considered in our work, we are limited to available online information. Consequently, we are unable to determine internal details such as the details of their storage systems.

\parhead{Conclusion and Construct Validity.}
\looseness=-1
None of the common threats to conclusion and construct validity provided by Wohlin et al.~\cite{Wohlin2012} applies to our study.

\section{Related Work}
\label{sec:related-work}



\noindent
\looseness=-1
There are currently a few numbers of existing surveys and comparisons of tools with asset management support. We expect more studies in the future as discussions on standardized ML asset management and applied SE engineering practices in ML development deepen.

\looseness=-1
Isdahl et al. \cite{Isdahl2019} surveyed ML platforms' support for reproducibility of empirical results. Several platforms considered in the study falls under the ML experiment management systems—which is also the focus of our study. The authors proposed a method to assess ML platforms' reproducibility and analyzed the features which improve their support. Ferenc et al. \cite{Ferenc2020} carried out a comparison of ML frameworks' features, investigating support for features that include data versioning, graphical dashboards, model versioning, and ML workflow support. Weißgerber et al. \cite{Weibgerber2019} investigate 40 ML open source platforms for support of full-stack ML research with open science at its core. The authors developed an open science-centered process model that integrates transparency and openness for ML research. The authors found 11 tools and platforms to be most central to the research process. They further analyzed them for resource management and model development capabilities.

\looseness=-1
Similar to our work, these previous studies have considered tools such as StudioML\xcite{studioml}, MLFlow\xcite{mlflowweb, Zaharia2018}, Weights and Biases\xcite{wand}, Polyaxon\xcite{polyaxon}, Comet.ml\xcite{cometml}, Sacred\xcite{Greff2017}, Sumatra\xcite{sumatra}, and DVC\xcite{dvc}. In contrast to our work, they\cite{Isdahl2019,Ferenc2020, Weibgerber2019} adopted a more coarse-grained understanding of assets and their management operation. This present work is the first systematic investigation of supported asset types (e.g., differentiating between models and data), which is an essential element of the ML domain and has practical implications to users of the considered tools (see the discussion in \cref{sec:discussion}).

\section{Conclusion}
\label{sec:conclusion}

\noindent
\looseness=-1
This paper discussed asset management as an essential discipline to scale the engineering of ML-based systems and of ML experiments. It also presented a survey of 17 systematically selected tools with management support for ML assets, identifying their common and distinguishing features. We focused on tools addressing commonly reported ML engineering challenges---mainly experiment management, monitoring, and logging. We performed a feature-based analysis and reported our findings using feature models. We identified five top-level features, namely,
supported asset types,
collection methods,
storage methods,
supported operations,
and integration
to other management tools. We found that over half of our subject tools depend on traditional version control systems for asset tracking, and the abstraction level for assets does not often distinguish between important ML asset types, such as models and datasets.

\looseness=-1
In future work, we intend to explore proposed ML asset management tools from the literature in addition to the ones used in practice. Such a study will provide an overview characterization of the solution space of the ML asset management techniques. We will provide a more elaborate representation that captures the dependencies between assets as observed in the tools.
To study usefulness and adequacy for addressing user needs, we plan to perform user studies, investigating the usage of functionalities provided by the tools.

\parhead{Acknowledgement.} Wallenberg Academy Sweden.


\bibliographystyle{IEEEtran}
\bibliography{doc}

\end{document}